# *Spectral Remittances of Chiral Sculptured Zirconia Thin Films in Non-axial propagation*

Ferydon Babaei[1,*] and Hadi Savaloni[2]
*1) Department of Physics, University of Qom, Qom, Iran.*
*2) Department of Physics, University of Tehran, North-Kargar Street, Tehran, Iran.*
**) Corresponding author: Tel: +98 251 2853311; Fax: +98 251 2854972; Email: fbabaei@qom.ac.ir*

## Abstract

The transmission and reflection spectra from a right-handed chiral sculptured zirconia thin film are calculated using the piecewise homogeneity approximation method and the Bruggeman homogenization formalism and considering that the dispersion of dielectric function happens in non-axial propagation state. First and second Bragg peaks were observed in cross-poloraized reflectances and it became clear that increasing the azimutal angle affects the spectra of linearly polarized state significantly. This is opposite to circularly polarized state.

## 1. Introduction

In obtaining the transmission and reflection spectra for sculptured thin films, the relative dielectric constant of the film material at a certain frequency is usually considered, and then Bruggeman homogenization formalism is used to estimate the relative permittivity scalars. These scalar quantities are assumed constant in the procedure of obtaining the transmission and reflection spectra at all frequencies [1-3]. However, it is clear that the relative dielectric constant is dependent on the frequency and at each frequency it has a particular value. In addition, in most of the papers on the dielectric dispersion function effect on remittances (reflection and transmission) of sculptured thin films, the simple single-resonance Lorentzian model is used [4-6], while when this model is used there exist some difficulties in obtaining the oscillator strengths, resonance wavelengths, and absorption linewidths.

In our earlier work [7] we repoted the reflectance and transmittance from an axially excited chiral sculptured zirconia thin film. The aim of this work is to report on the real dispersion effect in chiral sculptured zirconia thin film, by using the experimental data of relative dielectric constant of thin film at each frequency [8] and to investigate the influence of azimutal and polar angles on these spectra, in conjunction with the piecewise homogeneity approximation method and the Bruggeman homogenization formalism.

## 2. Theory

Assume that the space ( $0 \le z \le d$ ) is occupied by a chiral sculptured thin film and that this film is being excited by a plane wave which propagates with an angle $\theta_{inc}$ relative to z axis and angle $\psi_{inc}$ relative to x axis in xy plane. The phasors of incident, reflectance and transmittance electric fields are given by [9]:

$$\begin{cases} \underline{E}_{inc}(\underline{r}) = [(\frac{i\underline{s} - \underline{p}_{+}}{\sqrt{2}})a_L - (\frac{i\underline{s} + \underline{p}_{+}}{\sqrt{2}})a_R]e^{i\underline{K}_0 \cdot \underline{r}} \quad , z \le 0 \\ \underline{E}_{ref}(\underline{r}) = [-(\frac{i\underline{s} - \underline{p}_{-}}{\sqrt{2}})r_L + (\frac{i\underline{s} + \underline{p}_{-}}{\sqrt{2}})r_R]e^{-i\underline{K}_0 \cdot \underline{r}} \quad , z \le 0 \\ \underline{E}_{tr}(\underline{r}) = [(\frac{i\underline{s} - \underline{p}_{+}}{\sqrt{2}})t_L - (\frac{i\underline{s} + \underline{p}_{+}}{\sqrt{2}})t_R]e^{i\underline{K}_0 \cdot (\underline{r} - d\underline{u}_z)} \quad , z \ge d \end{cases} \tag{1}$$

The magnetic field’s phasor in any region can be obtained from the relation:

$$\underline{H}(\underline{r}) = (i\omega\mu_0)^{-1}\underline{\nabla} \times \underline{E}(\underline{r})$$

where $(a_L, a_R)$, $(r_L, r_R)$ and $(t_L, t_R)$ are the amplitudes of incident plane wave, and reflectance and transmittance waves with left- or right-handed polariztions. We also have;

$$\begin{cases} \underline{r} = x\underline{u}_x + y\underline{u}_y + z\underline{u}_z \\ \underline{K}_0 = K_0(\sin\theta_{inc}\cos\psi_{inc}\,\underline{u}_x + \sin\theta_{inc}\sin\psi_{inc}\,\underline{u}_y + \cos\theta_{inc}\,\underline{u}_z) \end{cases} \tag{2}$$

where $K_0 = \omega\sqrt{\mu_0\varepsilon_0} = 2\pi/\lambda_0$ is the free space wave number, $\lambda_0$ is the free space wavelength and $\varepsilon_0 = 8.854\times10^{-12}\,Fm^{-1}$ and $\mu_0 = 4\pi\times10^{-7}\ Hm^{-1}$ are the permittivity and permeability of free space (vacuum), respectively. The unit vectors for linear polarization parallel and normal to the incident plane, $\underline{s}$ and $\underline{p}$, respectively are defined as:

$$\begin{cases} \underline{s} = -\sin\psi_{inc}\,\underline{u}_x + \cos\psi_{inc}\,\underline{u}_y \\ \underline{p}_{\pm} = \mp(\cos\theta_{inc}\cos\psi_{inc}\,\underline{u}_x + \cos\theta_{inc}\sin\psi_{inc}\,\underline{u}_y) + \sin\theta_{inc}\,\underline{u}_z \end{cases} \tag{3}$$

and $\underline{u}_{x,y,z}$ are the unit vectors in Cartesian coordinates system.

The reflectance and transmittance amplitudes can be obtained, using the continuity of the tangential components of electrical and magnetic fields at two interfaces, $z = 0$ and $z = d$, and solving the algebraic matrix equation [9]:

$$\begin{bmatrix} i(t_L - t_R) \\ -(t_L + t_R) \\ 0 \\ 0 \end{bmatrix} = \left[\underline{\underline{K}}(\theta_{inc},\psi_{inc})\right]^{-1}.\left[\underline{\underline{B}}(d,\Omega)\right].\left[\underline{\underline{M}}'(d,\Omega,\kappa,\psi_{inc})\right].\left[\underline{\underline{K}}(\theta_{inc},\psi_{inc})\right].\begin{bmatrix} i(a_L - a_R) \\ -(a_L + a_R) \\ -i(r_L - r_R) \\ (r_L + r_R) \end{bmatrix} \tag{4}$$

and:

$$\left[\underline{\underline{K}}(\theta_{inc},\psi_{inc})\right] = \begin{bmatrix} -\sin\psi_{inc} & -\cos\psi_{inc}\cos\theta_{inc} & -\sin\psi_{inc} & \cos\psi_{inc}\cos\theta_{inc} \\ \cos\psi_{inc} & -\sin\psi_{inc}\cos\theta_{inc} & \cos\psi_{inc} & \sin\psi_{inc}\cos\theta_{inc} \\ -\eta_0^{-1}\cos\psi_{inc}\cos\theta_{inc} & \eta_0^{-1}\sin\psi_{inc} & \eta_0^{-1}\cos\psi_{inc}\cos\theta_{inc} & \eta_0^{-1}\sin\psi_{inc} \\ -\eta_0^{-1}\sin\psi_{inc}\cos\theta_{inc} & -\eta_0^{-1}\cos\psi_{inc} & \eta_0^{-1}\sin\psi_{inc}\cos\theta_{inc} & -\eta_0^{-1}\cos\psi_{inc} \end{bmatrix} \tag{5}$$

$$\left[\underline{\underline{B}}(z,\Omega)\right] = \begin{bmatrix} \cos(\pi z/\Omega) & -\sin(\pi z/\Omega) & 0 & 0 \\ \sin(\pi z/\Omega) & \cos(\pi z/\Omega) & 0 & 0 \\ 0 & 0 & \cos(\pi z/\Omega) & -\sin(\pi z/\Omega) \\ 0 & 0 & \sin(\pi z/\Omega) & \cos(\pi z/\Omega) \end{bmatrix} \tag{6}$$

$\eta_0 = \sqrt{\dfrac{\mu_0}{\varepsilon_0}}$ is the intrinsic impedance of vacuum. The 4 x 4 matrix $[\underline{\underline{M}}'(z,\Omega,\kappa,\psi_{inc})]$ satisfies the differential matrix equation [9]:

$$\frac{d}{dz}[\underline{\underline{M}}'(z,\Omega,\kappa,\psi_{inc})] = i[\underline{\underline{P}}'(z,\Omega,\kappa,\psi_{inc})].[\underline{\underline{M}}'(z,\Omega,\kappa,\psi_{inc})] \quad ,0<z<d \qquad (7)$$

where:

$$\left[\underline{\underline{P}}'(z,\Omega,\kappa,\psi_{inc})\right] = \begin{bmatrix} 0 & -i\dfrac{\pi}{\Omega} & 0 & \omega\mu_0 \\ i\dfrac{\pi}{\Omega} & 0 & -\omega\mu_0 & 0 \\ 0 & \omega\varepsilon_0\,\varepsilon_c & 0 & i\dfrac{\pi}{\Omega} \\ \dfrac{\omega\varepsilon_0\,\varepsilon_b}{\tau} & 0 & i\dfrac{\pi}{\Omega} & 0 \end{bmatrix}$$
$$+\begin{bmatrix} -A\cos\xi\sin 2\chi & 0 & -B\sin\xi\cos\xi & -B\cos^2\xi \\ A\sin\xi\sin 2\chi & 0 & B\sin^2\xi & B\sin\xi\cos\xi \\ C\sin\xi\cos\xi & C\cos^2\xi & 0 & 0 \\ -C\sin^2\xi & -C\sin\xi\cos\xi & -A\sin\xi\sin 2\chi & -A\cos\xi\sin 2\chi \end{bmatrix} \qquad (8)$$

where in these equations, $A = \dfrac{\kappa(\varepsilon_b - \varepsilon_{a)}}{2\varepsilon_a\tau}$ , $B = \dfrac{\kappa^2}{\omega\varepsilon_0\varepsilon_a\tau}$ , $C = \dfrac{\kappa^2}{\omega\mu_0}$,

$\xi = (\dfrac{\pi\, z}{\Omega}) - \psi_{inc}$ and $\tau = \cos^2\chi + (\varepsilon_b/\varepsilon_a)\sin^2\chi$ . $\kappa = K_0\sin\theta_{inc}$ and $\Omega$ is the half of the periodical structure and $\varepsilon_{a,b,c}$ are the relative permittivity scalars and $\chi$ is the rise angle for the CSTF.

In order to obtain $[\underline{\underline{M}}'(d,\Omega,\kappa,\psi_{inc})]$ the piecewise homogeneity approximation method is used. In this method the CSTF is divided into N (a big enough number) very thin layers with a thickness of $h = d/N$ (5 nm will suffice). Using the equality of the tangential components of fields at interfaces the matrix $[\underline{\underline{M}}'(d,\Omega,\kappa,\psi_{inc})]$ can be estimated [9]:

$$\left[\underline{\underline{M}}'(d,\Omega,\kappa,\psi_{inc})\right] \approx \left[\underline{\underline{M}}_{N-1}^{PH}\right].\left[\underline{\underline{M}}_{N-2}^{PH}\right].....\left[\underline{\underline{M}}_{1}^{PH}\right].\left[\underline{\underline{M}}_{0}^{PH}\right] \quad (9)$$

where:

$$\left\lfloor \underline{\underline{M}}_{j}^{PH} \right\rfloor = e^{\{ih[\underline{\underline{P}}'(z_{j+1/2},\Omega,\kappa,\psi_{inc})]\}}$$
$$= [\underline{\underline{I}}] + \lim_{r\to\infty}\sum_{n=1}^{r}\frac{(ih)^n[\underline{\underline{P}}'(z_{j+1/2},\Omega,\kappa,\psi_{inc})]^n}{n!} \quad (10)$$

The number of terms in Eq. (10) can be determined from two converging conditions [9]:

$$\begin{cases} \dfrac{\mathrm{Re}\{[(ih)^{r+1}[\underline{\underline{P}}'(z_{j+1/2},\Omega,\kappa,\psi_{inc})]^{r+1}/(r+1!)]_{\alpha\beta}\}}{\mathrm{Re}\{[\sum_{n=0}^{r}(ih)^n[\underline{\underline{P}}'(z_{j+1/2},\Omega,\kappa,\psi_{inc})]^n/n!]_{\alpha\beta}\}} \le 10^{-12} \\ \dfrac{\mathrm{Im}\{[(ih)^{r+1}[\underline{\underline{P}}'(z_{j+1/2},\Omega,\kappa,\psi_{inc})]^{r+1}/(r+1!)]_{\alpha\beta}\}}{\mathrm{Im}\{[\sum_{n=0}^{r}(ih)^n[\underline{\underline{P}}'(z_{j+1/2},\Omega,\kappa,\psi_{inc})]^n/n!]_{\alpha\beta}\}} \le 10^{-12} \end{cases}, 1\le\alpha,\beta\le 4 \quad (11)$$

when transmittance and reflectance amplitudes are obtained from Eq. (4), then we can obtain the reflectance and transmittance coefficients as:

$$\begin{cases} r_{i,j} = \dfrac{r_i}{a_j} \\ t_{i,j} = \dfrac{t_i}{a_j} \end{cases}, i,j = L,R \quad (12)$$

The reflectance and transmittance coefficients for planes waves of S and P polarization can easily be obtained [10]:

$$
\begin{cases}
r_{SS} = -\dfrac{(r_{LL} + r_{RR}) - (r_{LR} + r_{RL})}{2} \\
r_{SP} = i\,\dfrac{(r_{LL} - r_{RR}) + (r_{LR} - r_{RL})}{2} \\
r_{PS} = -i\,\dfrac{(r_{LL} - r_{RR}) - (r_{LR} - r_{RL})}{2} \\
r_{PP} = -\dfrac{(r_{LL} + r_{RR}) + (r_{LR} + r_{RL})}{2}
\end{cases}, \tag{13}
$$

$$
\begin{cases}
t_{SS} = \dfrac{(t_{LL} + t_{RR}) - (t_{LR} + t_{RL})}{2} \\
t_{SP} = -i\,\dfrac{(t_{LL} - t_{RR}) + (t_{LR} - t_{RL})}{2} \\
t_{PS} = i\,\dfrac{(t_{LL} - t_{RR}) - (t_{LR} - t_{RL})}{2} \\
t_{PP} = \dfrac{(t_{LL} + t_{RR}) + (t_{LR} + t_{RL})}{2}
\end{cases}, \tag{14}
$$

The transmittance and reflectance are obtained from:

$$
\begin{cases}
R_{i,j} = |r_{i,j}|^2 \\
T_{i,j} = |t_{i,j}|^2
\end{cases} ,i, j = L, R, S, P \tag{15}
$$

## 3. Numerical results and discussion

It was assumed that a right-handed zirconia sculptured thin film in its bulk state is formed, which occupies the space in the free space ($n = 1$) with a thickness $d$. In order to obtain the relative permittivity scalars we used the Bruggeman homogenization formalism. In this formalism, the film is considered as a two phase composite, vacuum phase and the material (inclusion) phase. These quantities are dependent on different parameters, namely, columnar form factor, fraction of vacuum phase (void fraction), the wavelength of free space and the refractive index of the films material (inclusion). It is considered that each column in a STF consists of a string of small and identical ellipsoids and are electrically small (i.e., small in a sense that their electrical interaction can be ignored). Therefore [7]:

$$\varepsilon_\sigma = \varepsilon_\sigma(\lambda_0, \varepsilon_s, f_v, \gamma_\tau^s, \gamma_b^s, \gamma_\tau^v, \gamma_b^v), \qquad \sigma = a, b, c \qquad (16)$$

where $f_v$ is the fraction of void phase, $\gamma_\tau^{s,v}$ is one half of the long axis of the inclusion and void ellipsoids, and $\gamma_b^{s,v}$ is one half of the small axis of the inclusion and void ellipsoids.

In all calculations the following parameters were fixed; $\gamma_b^s = 2$, $\gamma_\tau^s = 20$, $\gamma_b^v = 1\ \gamma_\tau^v = 1$, , $f_v = 0.6$, $\chi = 30^0$, $\Omega = 162nm$, $d = 40\Omega$, and a range of wavelengths $\lambda_0 \in (250nm - 850nm)$ was considered, where the refractive index varies from 2.64599 to 2.17282 for the lowest wavelength to highest wavelength, respectively [8]. On the basis of experimental results for the refractive index of zirconia in the wavelength region mentioned above the dielectric dispersion function was included in the results shown in Figs. 1-4.

In Fig. 1 the reflectance $(R_{LL}, R_{RL})$ and transmittance $(T_{LL}, T_{RL})$ spectra as a function of wavelength $\lambda_0$ for $\psi_{inc} = 0$ and different $\theta_{inc}$ are plotted when a plane wave LCP incidences on a right-handed zirconia CSTF. $R_{LL}$ increases with $\theta_{inc}$ for all wavelengths and at $\theta_{inc} = 90^0$ the reflectance becomes unity because the transmittance $T_{LL}$ becomes zero. It can be observed that there exist two peaks in each $R_{RL}$ plot (two corresponding troughs can be seen in $T_{RL}$ plots). These peaks are known as the first and the second Bragg peaks.

The reflectance $(R_{LR}, R_{RR})$ and transmittance $(T_{LR}, T_{RR})$ spectra as a function of wavelength $\lambda_0$ for $\psi_{inc} = 0$ and different $\theta_{inc}$ are plotted in Fig.2, when a plane wave RCP incidences on a right-handed zirconia CSTF. Since the structural direction of thin film is the same as the direction of polarization of the incident plane wave, the circular Bragg phenomena in each of $R_{RR}$ spectra and their corresponding trough in $T_{RR}$ plots can be clearly distinguished. There are two peaks (first and second Bragg peaks) in $R_{LR}$ plots similar to $R_{RL}$ and the corresponding troughs in $T_{LR}$ spectra. It can also be observed that the higher order peaks occur at smaller wavelengths, while their intensity is low and cannot be distinguished easily.

The effect of the linear polarizations (S or P) of the incidence plane wave on the Bragg peaks is shown in Figs. 3 and 4, respectively. In Fig.3. a S-polarized plane wave is considered to incidence on a right-handed zirconia CSTF, and the resultant reflectance $(R_{SS}, R_{PS})$ and transmittance $(T_{SS}, T_{PS})$ spectra are given as a function of wavelength $\lambda_0$ for $\psi_{inc} = 0$ and different $\theta_{inc}$. By increasing the $\theta_{inc}$ angle the $R_{SS}$ increases from zero until at $\theta_{inc} = 90^0$ it reaches to unity (this effect is reversed in $T_{SS}$ plots). In $R_{PS}$ spectra the first and second Bragg peaks and their corresponding valleys in $T_{PS}$ spectra are obvious at $\theta_{inc} < 90^0$. It can be seen that the first order peaks occur at longer wavelengths and in order to observe them one should extend the wavelength range above 850 nm. These peaks cannot be observed for $\theta_{inc} \le 30^0$ either. In Fig. 4 the reflectance $(R_{SP}, R_{PP})$ and transmittance $(T_{SP}, T_{PP})$ spectra as a function of wavelength $\lambda_0$ for $\psi_{inc} = 0$ and different $\theta_{inc}$ are plotted when a P-polarized plane wave incidences on a right-handed zirconia CSTF. Two peaks (first and second order Bragg peaks) and their corresponding trough can be seen in $R_{SP}$ and $T_{SP}$ spectra, respectively. In $R_{PP}$ plot the circular Bragg phenomenon and in $T_{PP}$ plot the corresponding troughs at $\theta_{inc} < 90^0$ appears (this is opposite to $R_{SS}$ ). Therefore, it can be deduced that when a P-polarized plane wave is incidence on a right-handed CSTF the interference is more constructive.

The reflectance $(R_{LL}, R_{RL})$ and transmittance $(T_{LL}, T_{RL})$ spectra as a function of $\theta_{inc}$ and $\psi_{inc}$ for a single wavelength of $\lambda_0 = 480nm$, when a plane wave LCP incidences on a right-handed zirconia CSTF are given in Fig. 5. From the $R_{LL}$ plot it can be deduced that increasing $\psi_{inc}$ has no significant influence on reflectance. However, increasing the $\theta_{inc}$ angle will change the value of reflectance from zero until it reaches unity at $\theta_{inc} = 90^0$ (apart from the fact that in the $60^0 \le \theta_{inc} \le 80^0$ region there exist some disorders in form of peaks and troughs). The situation is reversed in $T_{LL}$. Consecutive peaks and troughs can be observed in $R_{RL}$ plot by increasing, $\theta_{inc}$, while increasing $\psi_{inc}$ has no effect on the value of reflectance. Finally, the maximum value for $T_{RL}$ is about 0.8.

Fig. 6 shows the reflectance $(R_{LR}, R_{RR})$ and transmittance $(T_{LR}, T_{RR})$ spectra as a function of $\theta_{inc}$ and $\psi_{inc}$ for a single wavelength of $\lambda_0 = 480nm$, when a plane wave RCP incidences on a right-handed zirconia CSTF. The Consecutive peaks and troughs are obvious in $R_{LR}$ plot, but again increasing $\psi_{inc}$ has no effect on the value of reflectance and the maximum value for $T_{LR}$ is about 0.6. In $R_{RR}$ plot a trough can be seen in $10^0 \le \theta_{inc} \le 60^0$ region, while higher reflectance is obtained in other regions. In $T_{RR}$ plot the maximum reflectance is about 0.8 in the region $10^0 \le \theta_{inc} \le 30^0$ and at lower and higher $\theta_{inc}$ the reflectance decreases towards zero.

In Fig. 7 the reflectance $(R_{SS}, R_{PS})$ and transmittance $(T_{SS}, T_{PS})$ spectra as a function of $\theta_{inc}$ and $\psi_{inc}$ for a single wavelength of $\lambda_0 = 480nm$, when a S-polarized plane wave incidences on a right-handed zirconia CSTF are given. In this study case it can be observed that not only the consecutive peaks and troughs are observed by increasing the $\theta_{inc}$, but also for $\theta_{inc} \leq 20^0$ increasing $\psi_{inc}$ causes a significant increase in the intensity of reflectance, while for $\theta_{inc} > 20^0$ no effect can be observed. However, in case of $T_{SS}$ spectra, for $\theta_{inc} < 20^0$ increasing $\psi_{inc}$ shows a small effect, while for the other values of $\theta_{inc}$ variation of $\psi_{inc}$ has no effect on $T_{SS}$. In $R_{PS}$ and $T_{PS}$ plots, the peaks and troughs are indicative of the fact that variation of azimutal and polar angles have significant effects on the values obtained for $R_{PS}$ and $T_{PS}$ spectra.

Fig. 8 shows the reflectance $(R_{SP}, R_{PP})$ and transmittance $(T_{SP}, T_{PP})$ spectra as a function of $\theta_{inc}$ and $\psi_{inc}$ for a single wavelength of $\lambda_0 = 480nm$, when a P-polarized plane wave incidences on a right-handed zirconia CSTF. Again it can be observed that the peaks and troughs in the $R_{SP}$ and $T_{SP}$ plots, indicate that variation of azimutal and polar angles have significant effects on the values obtained for $R_{SP}$ and $T_{SP}$ spectra.

In $R_{PP}$ plot by increasing $\theta_{inc}$ not only the consecutive peaks and troughs are observed, but also for $\theta_{inc} \leq 20^0$ increasing $\psi_{inc}$ causes a significant increase in the intensity of reflectance, while for $\theta_{inc} > 20^0$ no effect can be observed. In case of $T_{PP}$ spectra, for $\theta_{inc} < 20^0$ increasing $\psi_{inc}$ shows a small effect, while for the other values of $\theta_{inc}$ variation of $\psi_{inc}$ has no effect on $T_{PP}$.

In summary, in this work we have been able to study the reflectance and transmittance spectra from a right-handed zirconia CSTF on the basis of experimental data for refractive index of zirconia as a function of wavelength in the 250 to 850 nm range, in circular and linear polarization states for non-axial propagation. In the spectra of cross-polarized reflectances two peaks (first and second Bragg peaks) and in cross-polarized transmittance spectra two corresponding troughs for each state of polarization are observed. It was also observed that in case of circular polarization increasing the azimutal angle has no particular effect on the reflectance and transmittance spectra obtained for a single wavelength of 480 nm. However, a significant effect was observed in case of linear polarization. In addition, it should be emphasize here that we did not use the simple single-resonance Lorantzian model which is usually used in order to describe the dispersion effect of relative dielectric permittivity scalar in a CSTF. This was done by using the experimental refractive index of zirconia thin film [8] in conjunction with the piecewise homogeneity approximation method and the Bruggeman homogenization formalism at each given frequency.

This is because to our opinion it is possible that all layers of the film are not supposed to obey from single-resonance Lorentzian model, but they may be subject to double-resonance [11] or multi-resonance [12,13].

## 4. Conclusions

the piecewise homogeneity approximation method and the Bruggeman homogenization formalism at each given frequency were used to compute the transmittance and reflectance of a right-handed zirconia CSTF for non-axial propagation. This was carried out by considering the refractive index of zirconia at each given frequency, individually, in the frequency range of 250 to 850 nm in the homogenization formalism. Therefore in this way dispersion of the dielectric function was introduced into our calculations. This method directly takes advantage from the experimental relative dielectric constant of thin film and avoids the use of simple dispersion model known as single-resonance Lorentzian model.

Two peaks (first and second Bragg peaks) in the spectra of cross-polarized reflectance and two corresponding troughs in the cross-polarized transmittance spectra for each state of polarization were obtained. It was also observed that in case of circular polarization increasing the azimutal angle has no particular effect on the reflectance and transmittance spectra obtained for a single wavelength of 480 nm. However, contrary to the circular polarization, a significant effect was observed in case of linear polarization.

**Acknowledgements**

This work was carried out with the support of the University of Tehran and the Iran National Science Foundation (INSF).

**Figure captions**

Figure 1. Reflectance and transmittance spectra as a function of wavelength $\lambda_0$ for $\psi_{inc} = 0$ and different $\theta_{inc}$, when a LCP plane wave incidences on a right-handed zirconia CSTF. a) reflectance; b) transmittance.

Figure 2. Reflectance and transmittance spectra as a function of wavelength $\lambda_0$ for $\psi_{inc} = 0$ and different $\theta_{inc}$, when a RCP plane wave incidences on a right-handed zirconia CSTF. a) reflectance; b) transmittance.

Figure 3. Reflectance and transmittance spectra as a function of wavelength $\lambda_0$ for $\psi_{inc} = 0$ and different $\theta_{inc}$, when a S-polarized plane wave incidences on a right-handed zirconia CSTF. a) reflectance; b) transmittance.

Figure 4. Reflectance and transmittance spectra as a function of wavelength $\lambda_0$ for $\psi_{inc} = 0$ and different $\theta_{inc}$, when a P-polarized plane wave incidences on a right-handed zirconia CSTF. a) reflectance; b) transmittance

Figure 5. Reflectance and transmittance spectra as a function of $\theta_{inc}$ and $\psi_{inc}$ for a single wavelength of 480 nm, when a LCP plane wave incidences on a right-handed zirconia CSTF. a) reflectance; b) transmittance

Figure 6. Reflectance and transmittance spectra as a function of $\theta_{inc}$ and $\psi_{inc}$ for a single wavelength of 480 nm, when a RCP plane wave incidences on a right-handed zirconia CSTF. a) reflectance; b) transmittance

Figure 7. Reflectance and transmittance spectra as a function of $\theta_{inc}$ and $\psi_{inc}$ for a single wavelength of 480 nm, when a S-polarized plane wave incidences on a right-handed zirconia CSTF. a) reflectance; b) transmittance

Figure 8. Reflectance and transmittance spectra as a function of $\theta_{inc}$ and $\psi_{inc}$ for a single wavelength of 480 nm, when a P-polarized plane wave incidences on a right-handed zirconia CSTF. a) reflectance; b) transmittance

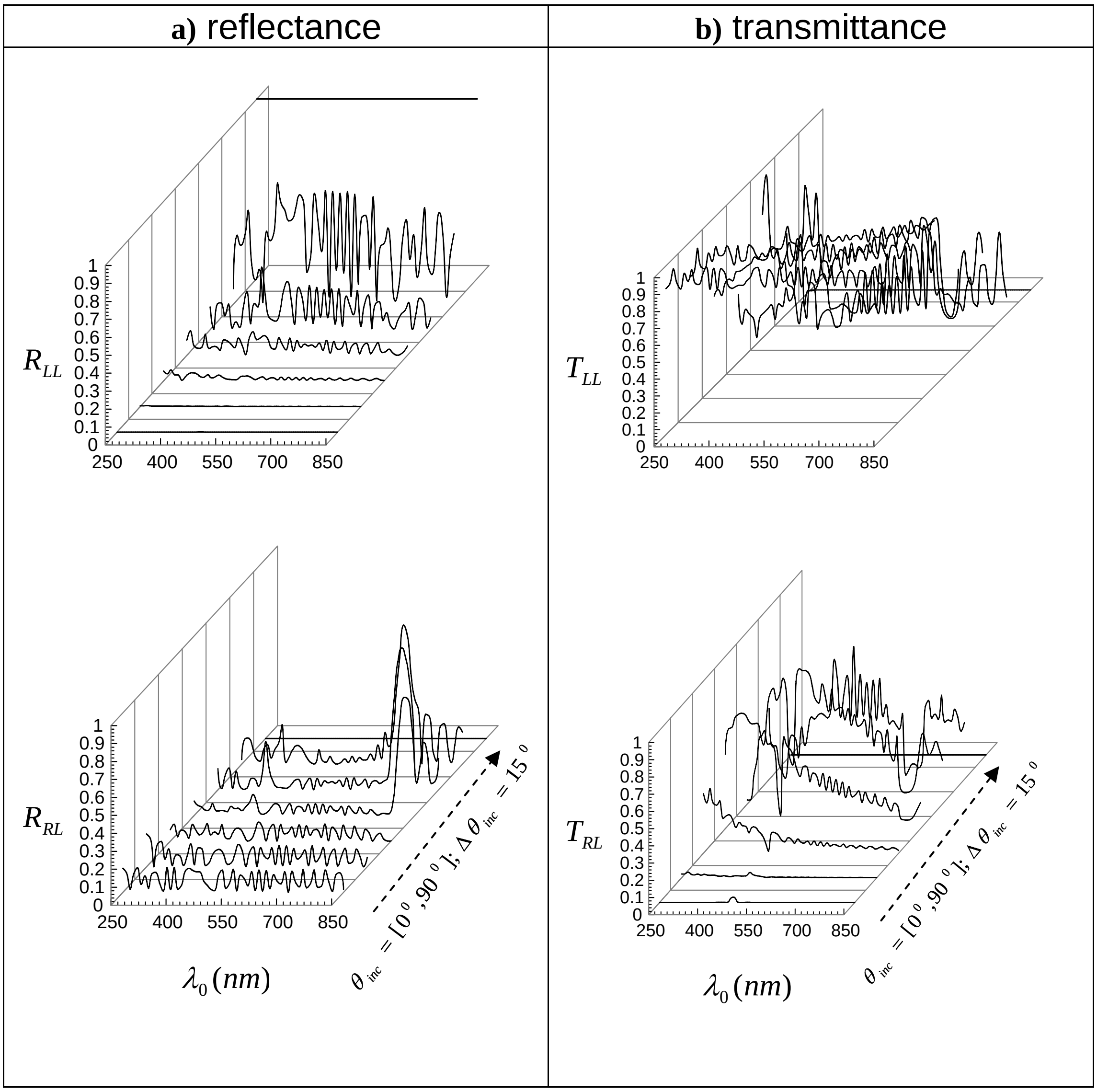

**Fig. 1; F. Babaei and H. Savaloni**

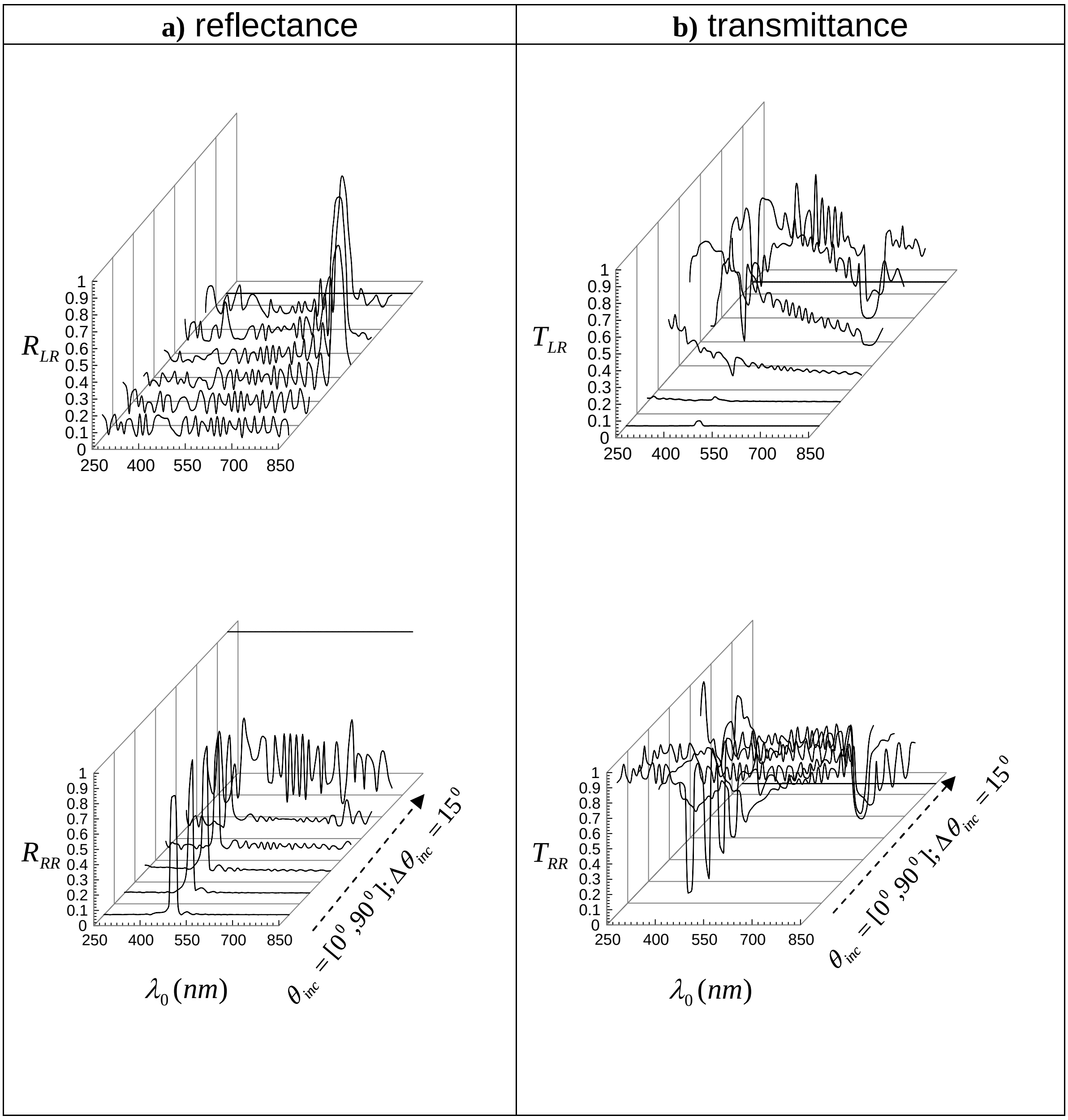

**Fig. 2; F. Babaei and H. Savaloni**

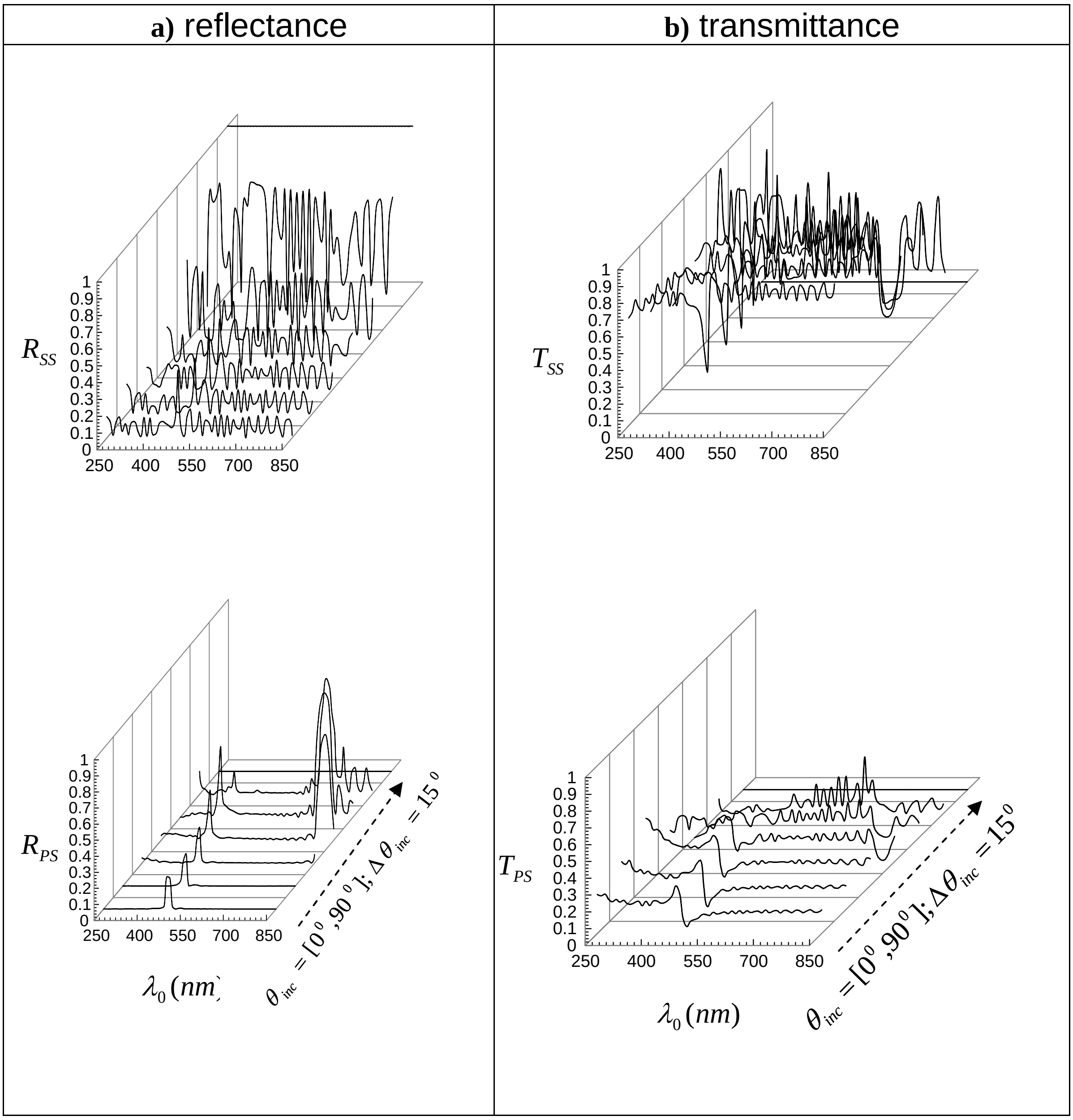

**Fig. 3; F. Babaei and H. Savaloni**

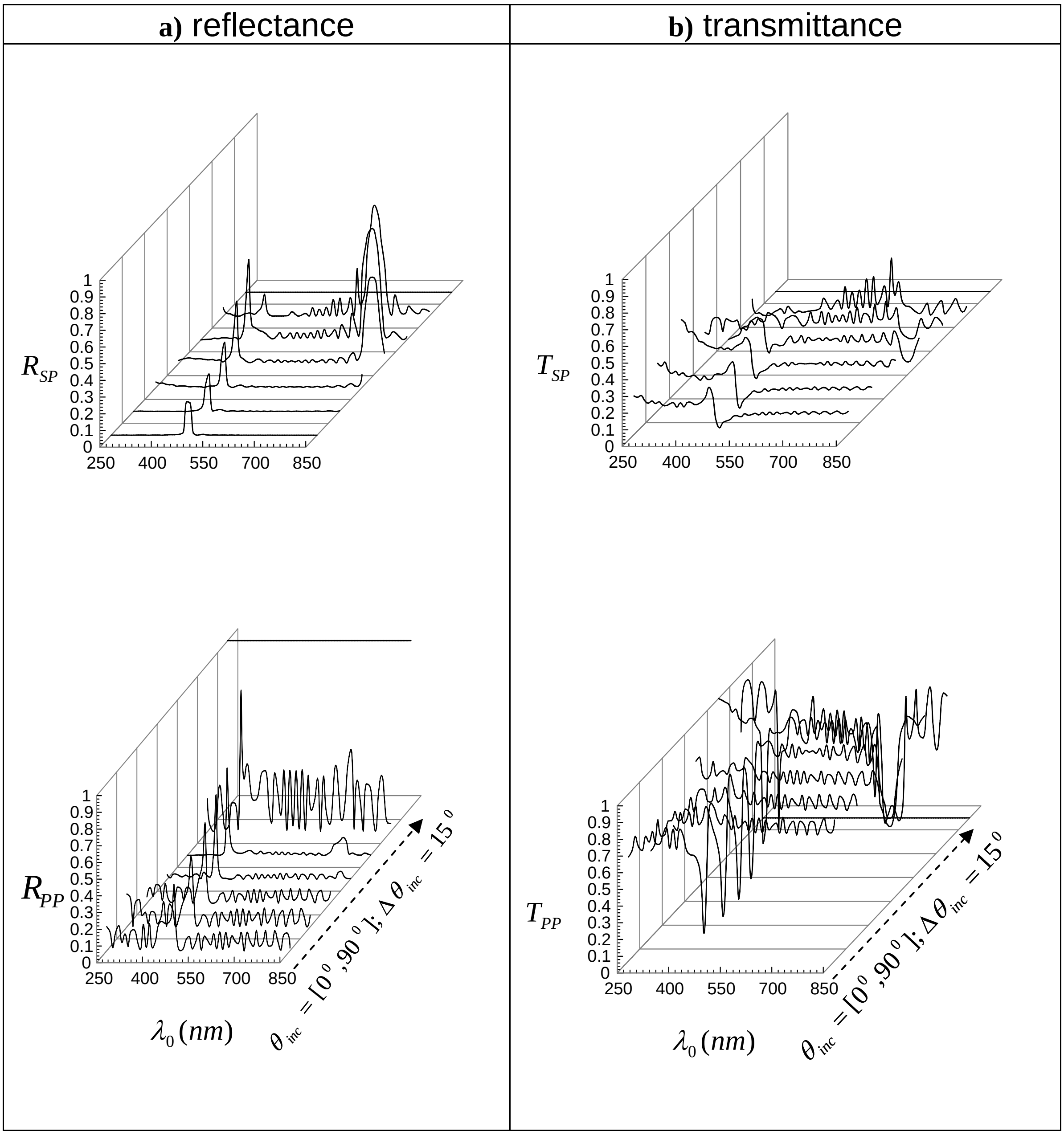

**Fig. 4; F. Babaei and H. Savaloni**

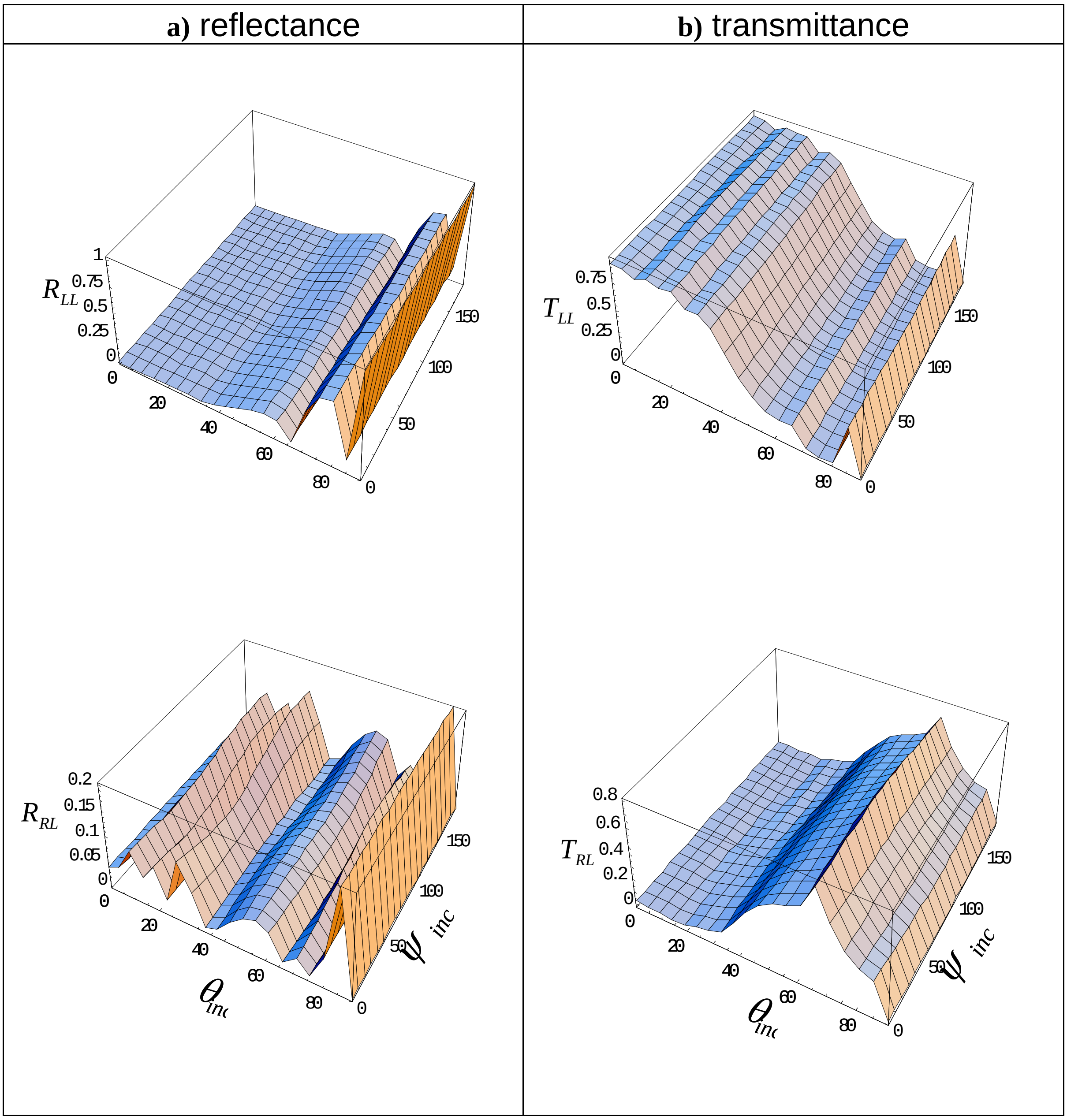

**Fig. 5; F. Babaei and H. Savaloni**

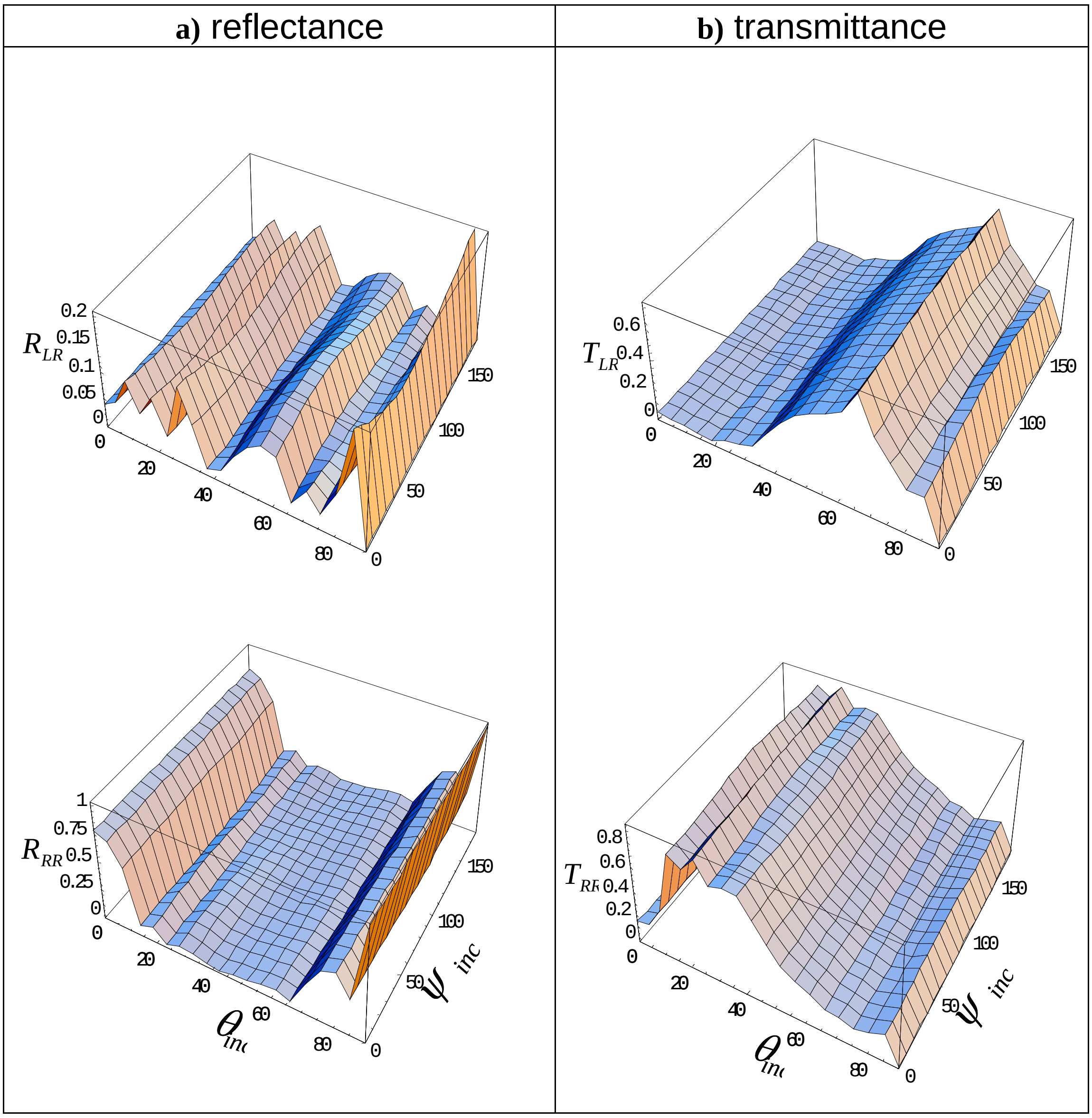

**Fig. 6; F. Babaei and H. Savaloni**

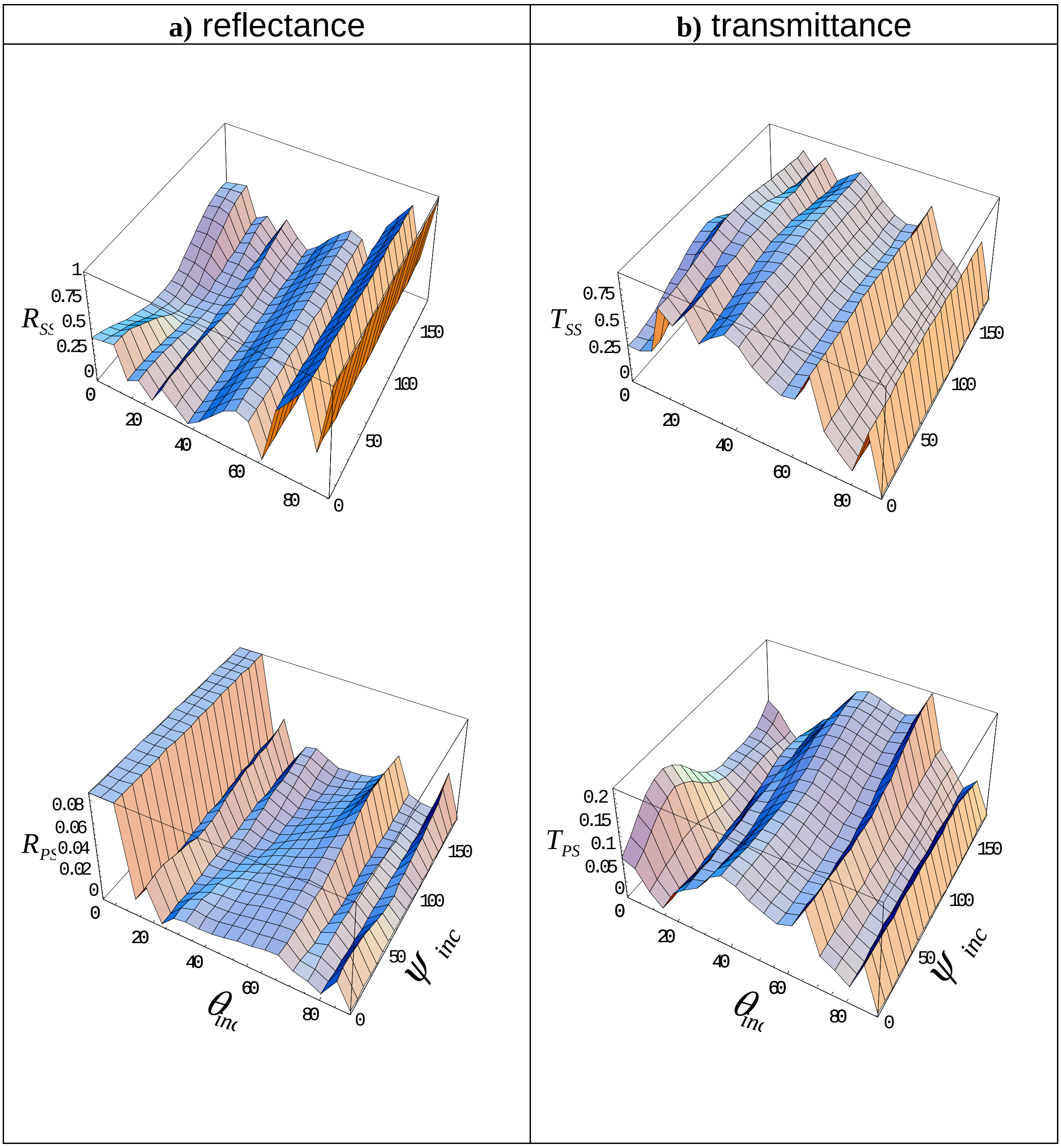

**Fig. 7; F. Babaei and H. Savaloni**

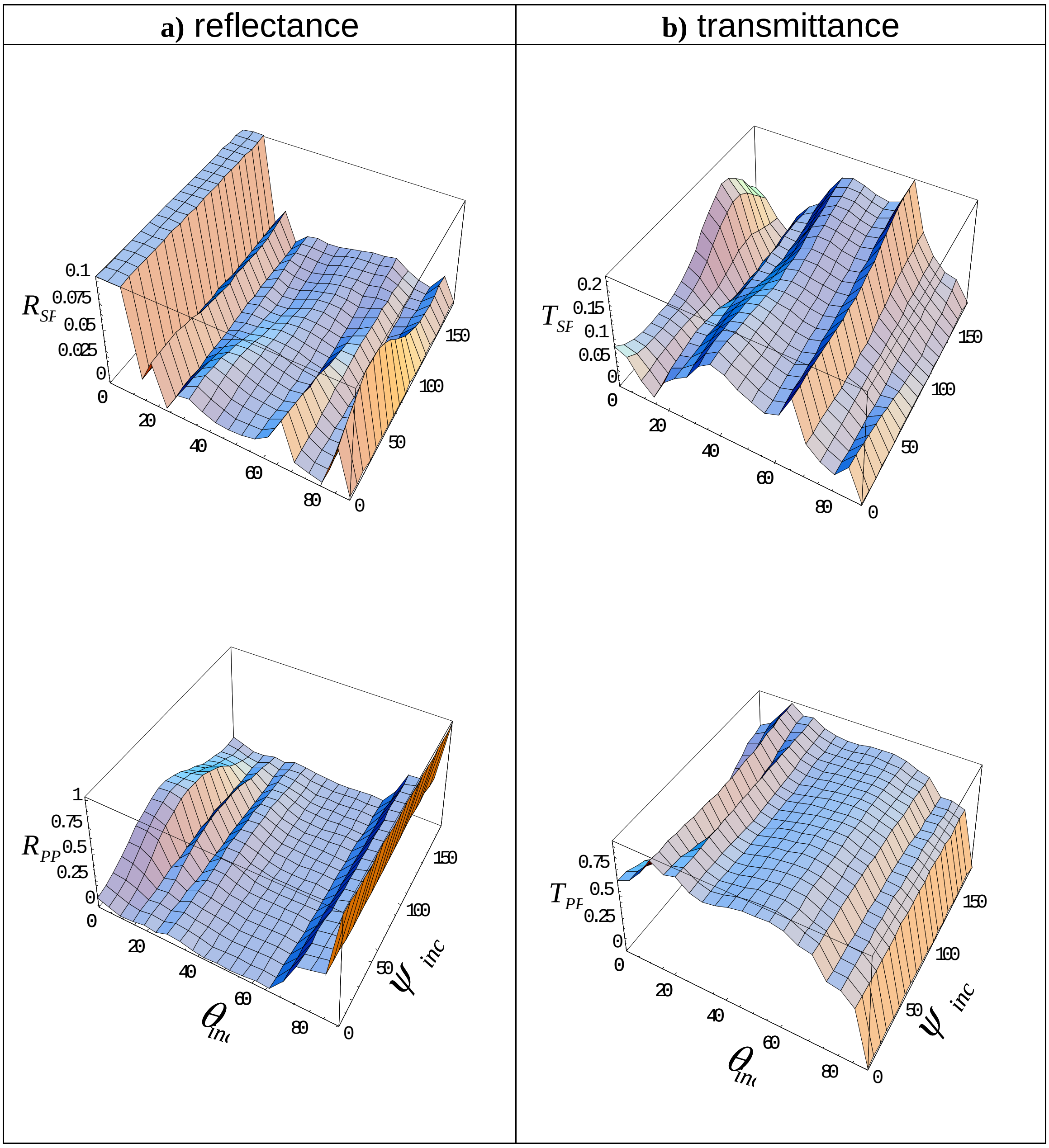

**Fig. 8; F. Babaei and H. Savaloni**